*Evidence for the Dominance of Indirect Effects in 50 Trophically-Based Ecosystem Networks*

Andria K. Salas[a] and Stuart R. Borrett[b]

[a]Corresponding author
601 S. College Road
Center for Marine Science and Dept. of Biology and Marine Biology
University of North Carolina at Wilmington
Wilmington, NC 28403
616.334.4864
aks2515@gmail.com

[b]601 S. College Road
Dept. of Biology and Marine Biology and Center for Marine Science
University of North Carolina at Wilmington
Wilmington, NC 28403
910.962.2411
borretts@uncw.edu

**Abstract**

Indirect effects are powerful influences in ecosystems that may maintain species diversity and alter apparent relationships between species in surprising ways. Here, we applied Network Environ Analysis to 50 empirically-based trophic ecosystem models to test the hypothesis that indirect flows dominate direct flows in ecosystem networks. Further, we used Monte Carlo based perturbations to investigate the robustness of these results to potential error in the underlying data. To explain our findings, we further investigated the importance of the microbial food web in recycling energy-matter using components of the Finn Cycling Index and analysis of Environ Centrality. We found that indirect flows dominate direct flows in 37/50 (74.0%) models. This increases to 31/35 (88.5%) models when we consider only models that have cycling structure and a representation of the microbial food web. The uncertainty analysis reveals that there is less error in the *I/D* values than the ±5% error introduced into the models, suggesting the results are robust to uncertainty. Our results show that the microbial food web mediates a substantial percentage of cycling in some systems (median = 30.2%), but its role is highly variable in these models, in agreement with the literature. Our results, combined with previous work, strongly suggest that indirect effects are dominant components of activity in ecosystems.

## 1 Introduction

What is an example of an isolated individual? Given that metabolism and reproduction are two criteria for living entities, no organism can be isolated (Jørgensen et al., 1992, 1999; Wilkinsin, 2006). Heterotrophs must consume carbon sources and autotrophs utilize nutrients within the soil or water that were made available by other living things. Sexual reproduction typically requires another individual of the same species. Thus, at a fundamental level, organisms must interact with their environment to grow, reach maturity, and reproduce. From these most basic interactions necessary for life develops the complex ecological community. In this paper we elucidated a part of ecosystem complexity by testing the generality of the systems ecology hypothesis that indirect effects tend to dominate direct effects in ecosystem interaction networks.

Indirect effects can have powerful influences on ecosystem functioning (e.g. Abrams et al., 1996; Higashi and Patten, 1989; Menge, 1997; Patten, 1983; Wootton, 1993, 1994; Yodzis, 1988). For example, they have been found to help maintain diversity (Bever, 1999, 2002) and contribute to the success of a particular habitat patch (Diekötter et al., 2007). Indirect effects also can produce surprising relationships between species (Bondavalli and Ulanowicz, 1999), and can result in overall positive interactions despite local negative direct interactions (Fath and Patten, 1998; Patten, 1991; Ulanowicz and Puccia, 1990). In addition, indirect effects can develop rapidly in ecosystems (Borrett et al., 2010; Menge, 1997). Given these diverse impacts, it is clear that understanding indirect effects is necessary to understand the processes that create, constrain, and sustain ecosystems.

This understanding is challenging to achieve given the difficulty in empirically studying indirect effects, necessitating the need for alternative approaches such as Network Environ Analysis (NEA). NEA is applied to ecosystem models of energy-matter flow and can be used to study indirect flows of energy (e.g. Fath and Borrett, 2006; Fath and Patten, 1999; Patten et al., 1976). The ecosystem network models we used in this work are simplified representations of ecosystems that trace the path of energy–matter flux through the system. In these models, nodes represent species, groups of species, or abiotic resource pools while the directed links represent the flow of energy–matter. NEA lets us partition these flows into direct and indirect components. Indirect flows are those in which a species receives energy indirectly from another species, such as when a polar bear receives energy from krill by consuming penguins that directly ate the krill. These indirect flows were the basis of our investigations.

We used NEA to test the hypothesis that the amount of energy flowing over indirect pathways tends to exceed the amount of energy flowing over direct pathways in ecosystems, a hypothesis that emerged in early applications of NEA (Patten, 1983, 1984, 1985). Higashi and Patten (1986, 1989) argued that the dominance of indirect flows over direct flows should be a general phenomenon by showing algebraically that indirect flow tends to increase with system size, connectance, cycling, and magnitude of direct flows. The hypotheses concerning system size and cycling are further supported by an analysis of large hypothetical ecosystem models built from a community assembly algorithm (Fath, 2004). These studies suggest that the dominance of indirect flows has the potential to occur in real ecosystems, but few studies report this phenomenon in empirically derived models (but see e.g. Borrett et al., 2006; Patten, 1983). Thus, the generality of the hypothesis in models of real systems remains to be tested. Further, little is known regarding the sensitivity of the results to uncertainty in the data used for network model construction (e.g., Borrett and Osidele, 2009; Kaufmann and Borrett, 2010).

There are many different types of indirect effects (Strauss, 1991 and Wootton, 1993; 1994), and NEA can capture those that are and are affected by energy–matter transactions. Indirect flows are a subset of indirect effects, but the former has the ability to reflect a wider range of indirect effects than feeding relationships. Indirect effects have the ability to increase, decrease, or have no affect on indirect flows. For example, a meta-analysis by Preisser et al. (2005) suggests that the act of intimidation of predators on prey results in reduced resource acquisition by the prey species. The larger the intimidation effect of predators on prey, the smaller the flows of energy to the prey species. While NEA does not capture all types of indirect effects, the consequences of many are reflected in energy transfers. Therefore, by comparing the intensity of indirect flows to direct flows, we are estimating the magnitude of indirect effects. More importantly, this estimation should tend to be conservative given the possibility of indirect effects that minimize the strength of indirect flows.

To understand the variability of the relative importance of indirect effects, we further investigated the species essential to energy–matter cycling. The microbial food web plays an important role in mediating energy–matter cycling. Much of the carbon fixed by primary producers is released as particulate and dissolved organic matter. Dissolved organic carbon (DOC) is a significant source of carbon for bacteria (Larsson and Hagström, 1982) and it is this relationship that is largely the basis of the microbial food web (Hart et al., 2000), which includes benthic and pelagic bacteria, ciliates, flagellates, and microzooplankton. Particulate organic carbon (POC) from plant material and animal wastes is also a medium for microbial colonization (Fenchel, 1970). The functioning of the microbial web and loop can reintroduce the energy-matter in DOC and other detrital groups to the larger metazoan food web (Azam et al., 1983; Pomeroy, 1974). Microzooplankton also are important to the flow of energy in the system as they make available the energy in nano- and picoplankton to larger consumers (Sherr and Sherr, 1988). The importance of detritus and the microbial food web has been recognized for a variety of aquatic systems, from rivers to lakes to coastal systems (Mann, 1988). However, the amount of primary production that actually passes through the microbial food web as a conduit to the metazoan food web is highly variable (as in Hart et al., 2000) and system dependent (Hart et al., 2000; Sommer et al., 2002). In summary, the microbial food web should be considered when investigating the pathways of carbon flow given its importance to the recycling process.

In this work we specifically tested four hypotheses. First, we tested the generality of the dominance of indirect effects hypothesis in a set of 50 empirically-derived ecosystem models. Second, we tested the idea that indirect flows through an ecosystem network tend to increase with system size, connectance, cycling, and direct flow. Third, we tested the robustness of the analytical results to potential data uncertainty in the network models using an uncertainty analysis. Finally, given the generally important role of the microbial food web in ecosystem carbon flow, we tested the hypothesis that the representation and importance of this subweb in the models helps explain why indirect flows do or do not dominate in these ecosystems.

# 2 Methods and Materials

## 2.1 Model Database

To test the hypothesis that indirect flows are dominant in ecosystems, we analyzed 50 ecosystem models using NEA. To build our database of empirically derived ecosystem models, we started with Ulanowicz's data set (http://www.cbl.umces.edu/~ulan/) and then continued to search for models published in the literature. For this study, we selected ecosystems models that focused on trophic interactions that typically trace the movement of carbon or energy. We excluded models that might be considered more biogeochemically based (see Christian, et al. 1996; Baird, et al. 2008; and Borrett, 2010 for more on this distinction). We also selected models that were at least minimally empirically based. These models were designed for a specific system and parameterized with some information gathered from the field. However, some literature values from related systems or best approximations may also have been used. Thus, the models range in the extent to which they were empirically derived, as some are largely field based and would be considered more empirical than those that are based solely on literature. Table 1 lists the ecosystem network models used in this study and describes some of the key features of the models.

This collection of 50 network models represents 35 distinct mostly freshwater and marine systems. Some of the models represent the same system at different times of the year or under different ecological conditions. Our database includes 18 of 48 models found in Ulanowicz's collection. Many of the models were initially reported at steady-state (inputs = outputs), but we balanced 15/50 of the models using the AVG2 algorithm (Allesina and Bondavalli, 2003) to meet the assumptions of NEA (see next section). These are the same models used by Borrett and Salas (2010) to test the resource homogenization systems ecology hypotheses.

## *2.2* Network Environ Analysis

### 2.2.1 Network Model Information

We applied NEA to all models using a modified version of the NEA.m MATLAB® function (Fath and Borrett, 2006). Here we introduce the analysis using a hypothetical example model of an oversimplified ecosystem network consisting of detritus, bacteria, amphipods, and blue crabs (Figure 1a). In this example, the number of nodes, $n$, in the network is 4. The adjacency matrix ($\mathbf{A}_{n \times n} = a_{ij}$) captures the structural information of the network. When there is a flow from $j$ to $i$, $a_{ij} = 1$, and $a_{ij} = 0$ otherwise. Note that NEA uses a column to row orientation. In our example, bacteria decompose detritus, resulting in a 1 in the matrix at $a_{21}$. The amount of energy-matter flowing between the nodes is represented in the $\mathbf{F}_{n \times n} = f_{ij}$ matrix (Figure 1a), in which the 1's of the adjacency matrix are replaced by the estimated flow values with general units of M $L^{-2 \text{ or } -3}$ $T^{-1}$ in mass, length, and time dimensions. Transfer of energy across the system boundary is represented in $\vec{z}$ and $\vec{y}$, which are vectors of the boundary inputs and outputs, respectively (Figure 1a). The former includes processes such as immigration or transport of organic material into the system. The latter includes processes such as fishing, emigration, and transport of organic material out of the system. Also included in $\vec{y}$ is respiration from all living nodes. With this system information, we can begin our analysis.

#### 2.2.2 Throughflow Analysis

Node throughflow is the amount of energy or matter that flows through each node. This can be calculated from an input $T_i^{in} = \sum_{j=1}^{n} f_{ij} + z_i$ or output $T_j^{out} = \sum_{i=1}^{n} f_{ij} + y_j$ perspective. $T^{in}$ sums all flows into node $i$ from nodes $j$, and adds to this quantity the boundary inputs into node $i$, while $T^{out}$ sums all flows that exit node $j$ to go to nodes $i$, in addition to the boundary outputs from node $j$. At steady state, $T^{in} = T^{out} = T = (T_j,)$ where $T_j$ is the total amount of energy entering or exiting each node $j$. Total system throughflow ($TST = \sum T_j$) is the total amount of energy or matter that flows through the system, and its meaning can be likened to a gross domestic product (GDP). GDP is the total monetary flow of a country's economy, and TST is the total substance flow in an ecosystem.

Next, we determine the direct flow intensity, $\mathbf{G} = (g_{ij})$, between any two nodes. Using an output perspective, we divide the values in each row $i$ of the **F** matrix by node throughflow, $\mathbf{G} = (g_{ij}) = f_{ij} / T_j$. Thus, $g_{ij}$ is the percentage of each node's throughflow that goes to each node with which it directly interacts. For example, in Figure 1b 33% of bacteria's energy flows to detritus.

Both **A** and **G** can be raised to exponents to examine indirect pathways and flow intensities. The matrix $\mathbf{A}^2$, for example, gives the numbers of pathways of length two ($m = 2$) that exist between all node pairs. The matrix $\mathbf{G}^2$ gives the flow intensities between the node pairs over these pathways of length two. For example, there is one indirect pathway of length 2 from amphipods to bacteria (via detritus), and 33% of the amphipod's energy flows to bacteria via this indirect pathway (Figure 1b). The same interpretation can be given to $\mathbf{A}^m$ and $\mathbf{G}^m$ for any $m$. Thus, the flow intensity between the nodes in the network over all indirect path lengths ($m > 2$) is equal to $\sum_{m=2}^{\infty} \mathbf{G}^m$.

If the system graph is a single strongly connected component such that each node is reachable from every other node by a pathway of some length, then the $\mathbf{G}^m$ matrix will be entirely composed of non-zero numbers by some value of $m$ (see Borrett et al., 2007, 2010). This contrasts with the direct flow intensity matrix **G**, which in almost all cases will have many zeros as it is unlikely there are direct connections between all nodes. An analysis of pathways through the **A** and **G** matrices with increasing $m$ can be found in more detail in Patten (1985). In our example (Figure 1b), we see that $\mathbf{G}^2$ has 8 zeros, versus 10 in the **G** matrix, and $\mathbf{G}^{10}$ has no zero values. This is significant because it reveals that all nodes are connected indirectly via indirect paths. Even though the quantity of energy flowing over paths of increasing length tends to decrease due to dissipation, longer paths are still being utilized and may significantly contribute to TST (Borrett et al. 2010). Each node receives some contribution to its throughflow from each and every other node from which it is reachable, further suggesting that indirect flows are an important consideration. The $\mathbf{G}^m$ matrix series can be used to determine the integral or total flow intensity, which is captured in the matrix **N**:

$$\mathbf{N} \equiv \sum_{m=0}^{\infty} \mathbf{G}^m = \underbrace{\mathbf{G}^0}_{Boundary} + \underbrace{\mathbf{G}^1}_{Direct} + \underbrace{\mathbf{G}^2 + \mathbf{G}^3 + \ldots + \mathbf{G}^m + \ldots}_{Indirect}. \quad (1)$$

The matrix identity, $\mathbf{G}^0 = \mathbf{I}$, is an *n x n* matrix of zeros with ones along the principle diagonal and is used to map in the boundary inputs, $\mathbf{G}^1$ is the direct flow intensity, and $\mathbf{G}^m$ is the indirect flow intensity over paths of length *m*. Each element in **N** represents the total flow intensity between the nodes represented by that element's row and column. Multiplication by $\vec{z}$ generates the realized flow values (Borrett et al. 2006; Borrett and Freeze, in press). These values can be summed to return the total system throughflow as

$$\begin{aligned}\sum T &= \sum \mathbf{N}\vec{z} \\ &= \underbrace{\sum \mathbf{G}^0\vec{z}}_{Boundary} + \underbrace{\sum \mathbf{G}^1\vec{z}}_{Direct} + \underbrace{\sum (\mathbf{G}^2 + \mathbf{G}^3 + \ldots + \mathbf{G}^m \ldots)\vec{z}}_{Indirect} \cdot \end{aligned} \quad (2)$$

The power series in equation 1 should be convergent because ecosystems are thermodynamically open (Jørgensen et al., 1999) and energy-matter is lost at each transfer. Thus, we can determine **N** exactly using the identity $\mathbf{N} = (\mathbf{I} - \mathbf{G})^{-1}$.

We can use equation 2 to determine the amount of indirect flow relative to direct flow as

$$I/D = \frac{\sum \mathbf{N}\vec{z} - \sum \mathbf{G}\vec{z} - \sum \mathbf{I}\vec{z}}{\sum \mathbf{G}\vec{z}}. \quad (3)$$

If the *I/D* ratio is greater than one, it indicates that the magnitude of indirect flow is greater than direct flow. This would be evidence to support the dominance of indirect effects hypothesis.

If we divide both sides of equation (2) by $\sum T$, we obtain the relative proportions of TST derived from boundary, direct, and indirect flows. Indirect flow intensity (IFI) is the proportion of total system throughflow derived from indirect pathways and direct flow intensity (DFI) is the proportion of total system throughflow due to direct pathways. Specifically, IFI and DFI are calculated as

$$\text{IFI} = \frac{\sum ((\mathbf{N} - \mathbf{I} - \mathbf{G})\vec{z})}{\sum T}, \text{ and} \quad (4)$$

$$\text{DFI} = \frac{\sum (\mathbf{G}\vec{z})}{\sum T}. \quad (5)$$

To test the generality of the dominance of indirect flows (hypothesis 1), we calculated *I/D* for the models in our data set. Further, we used a linear regression and Spearman's rank correlation test to test the hypothesized relationship between IFI and recycling, DFI, system size, and connectance ($C = L / n^2$, where $L$ is the number of links in the network).

**2.3 Contributions of nodes to system flux**

To test the hypothesized importance of the microbial food web in energy–matter cycling (hypothesis 4), we used two approaches: a partitioning of recycled matter into the fraction derived from the microbial food web and other compartments, and Environ Centrality (EC).

### 2.3.1 *Recycling*

In contrast to the decomposition of TST described in section 2.2.2, Finn (1976) introduced an alternative decomposition. He distinguished between flow from cyclic and acyclic pathways: $TST = cyclic + acyclic$. Cyclic flow is determined as:

$$Cyclic = \sum_{j=1}^{n} \left(\frac{n_{jj}-1}{n_{jj}}\right) T_j \qquad (6)$$

He then constructed what is known as the Finn Cycling Index (FCI) as: $FCI = cyclic/TST$.

Here, we subsetted the FCI to investigate the contribution of microbial food web to system recycling. We first identified the nodes in each model that represent microbial organisms. For convenience, we will conceptually reorder our network so that the first 1, 2, …, $k$ compartments are associated with the microbial food web. Then, we can partition the traditional FCI (Finn 1976; 1980) into the fraction mediated by the microbial food web (MFW) and the remainder of the nodes (other):

$$FCI = FCI_{MFW} + FCI_{other} \ . \qquad (7)$$

Therefore, the percentage of the system's cycling that is mediated by the microbial food web can be calculated by:

$$\begin{aligned} \%FCI_{MFW} &= \frac{FCI_{MFW}}{FCI} \\ &= \frac{\sum_{j=1}^{k} \left(\frac{n_{jj}-1}{n_{jj}} * T_j\right)}{\sum_{j=k+1}^{n} \left(\frac{n_{jj}-1}{n_{jj}} * T_j\right)} \end{aligned} \qquad (8)$$

### 2.3.2 *Environ Centrality*

To further characterize the role of specific nodes, selected network models were analyzed using environ centrality (EC; Fann, 2009). Centrality is a social science concept that "describes the location of individuals in terms of how close they are to the center of action in the network" (Hanneman and Riddle, 2005). Borgatti and Everett (2006) and Wasserman and Faust (1994) provide more detail on this concept. Fann (2009) introduced an environ based centrality metric that determines the role of each node in generating the total system throughflow. Following Estrada (2010), it is a globally weighted walks-based metric most similar to eigenvector centrality. This metric can be calculated for the input (*$EC^{in}$*) or output perspective (*$EC^{out}$*), and the average of the two orientations was used here (average environ centrality, *AEC*).

$$EC^{in} = \frac{\sum_{j=1}^{n} n_{ij}}{\sum_{i=1}^{n} \sum_{j=1}^{n} n_{ij}} \qquad (9)$$

$$EC^{out} = \frac{\sum_{i=1}^{n} n_{ij}}{\sum_{i=1}^{n} \sum_{j=1}^{n} n_{ij}} \qquad (10)$$

$$AEC = \frac{EC_j^{in} + EC_j^{out}}{2}. \qquad (11)$$

Thus, *AEC* gives the average proportional contribution that each node is making to total flow intensity. This metric indicates the relative significance of the nodes in generating system activity. We calculated this metric for the St. Marks seagrass system to better explain our results.

### 2.4 Analytical Robustness

We examined the reliability of our analytical results in two distinct ways; one addresses the potential problems caused by decisions made in the construction of the models and the other addresses the reliability of the models themselves to be reasonable representations of the underlying systems they represent.

#### *2.4.1 Model Structure*

Ecosystem network models range in their ability to reflect the actual underlying system and general overall quality. We initially choose to include all models available that have an empirical basis because the dominance of indirect flows is hypothesized to hold true for most networks. Yet all the models do not include the characteristics necessary to represent carbon flow that we believe are critical to be present for our work. Therefore, we did two analyses, one on the set of all models (Model Set 1) and another on those that meet more stringent criteria (Model Set 2). These criteria for Model Set 2 were that the model must 1) have network architecture that allows cycling, and 2) include species that are critical in mediating the process of energy cycling. These characteristics are necessary because cycles are known to be important components of ecosystem structure and function (Azam et al. 1986; Fath and Halnes 2007; Patten 1985). Species that must be included for adequate representation of the process of energy cycling include those of the microbial food web, as previously discussed. Model Set 2 is composed of 35 models representing 23 distinct systems.

#### *2.4.2 Uncertainty Analysis*

We tested the uncertainty in the *I/D* flow ratio to potential error in the model flow values (hypothesis 3) using a Monte Carlo procedure. Borrett and Osidele (2007) hypothesized that these results should be robust; here we tested this hypothesis. We constructed 10,000 models in which the original flow values were perturbed by ± 5%. We used the outputs for each node to rebalance the models by either subtracting or adding the amount of the change in the inputs of the node to that node's boundary output. This procedure allows the architecture, inputs, and the general distributions of flows for each network model to be maintained. Any modified model with a negative output was treated as a failed model because a negative loss does not make thermodynamic sense. We generated 10,000 successful models for each network model and applied NEA on each to estimate the analysis uncertainty.

## 3 Results

Our results generally confirm that indirect flows tend to dominate direct flows in empirically based networks. We present the evidence for each specific hypothesis in turn.

### 3.1 Dominance of Indirect Flows

Indirect flows dominate direct flows in 37 models out of 50 (74%) when the entire set of networks was considered (Figure 2, Model Set 1, all bars). When only models that met our more stringent set of criteria were considered, 31 models out of 35 (88.5%) show dominance of indirect flows (Figure 2, Model Set 2, black bars only). In Model Set 2, the St. Marks seagrass sites 3 and 4 and Cypress wet and dry seasons have an *I/D* <1.

Each of the 50 models in our data set are not necessarily independent given that some of the models reflect the same system, but under different ecological conditions or times of the year. However, they are not statistical replicates either. Given this ambiguity in the number of independent samples within our dataset, we also considered the results with respect to the number of distinct systems as opposed to models. When we do so, 27/35 systems (77.1%) exhibit dominance of indirect flows when all systems are considered. When only those systems that are used in Model Set 2 are analyzed this way, 21/23 systems (91.3%) support the hypothesis. Not included in the numerator of these calculations is the St. Marks Seagrass system, which has sites that both do and do not exhibit dominance of indirect flows.

### 3.2 Model Features Influencing Indirect Flows

System size $n$ ($r^2 = 0.069$; p = 0.06) and connectance $C$ ($r^2 = 0.022$; p = 0.31) do not explain much variation in IFI (Figure 3a and 3b, respectively). This is supported by Spearman's rank correlation test, as results were non-significant between IFI and both $n$ and $C$ (p = 0.078; ρ = -0.25 and p = 0.34; ρ = 0.14, respectively). In contrast, cycling does appear to have a positive nonlinear relationship with the magnitude of IFI (Figure 3d). We did not do a linear regression analysis of the relationship between DFI and FCI on IFI because these indicators are not independent calculations, violating the assumptions of the statistical method. However, the results of Spearman's rank correlation test show a significant correlation between IFI and both DFI and FCI ($p < 0.05$; ρ = -0.67 and $p < 0.05$; ρ = 0.88, respectively).

### 3.3 Analytical Robustness

For our uncertainty analysis, we randomly added ±5% (uniform random distribution) to the observed flows. In 34 out of 50 models we generated failed perturbed models because when balancing the model one or more adjusted values became negative. The number of failures ranged from 1,256 for Ythan Estuary to 464,485 for Cypress dry season. We found it impossible to generate 10,000 models in a reasonable time period with ±5% error for the Florida Bay wet and dry seasons. As these models have 14 zeros in the original output vector, we were unable to get 10,000 perturbed models that met our constraints. After generating the perturbed models, we verified that the maximum error was 5%, minimum error –5%, and median and mean error ≈ 0.

The error bars on Figure 2 show the full range of *I/D* uncertainties generated by our analysis, but here we will focus on the interquartile ranges, as this is more representative. The interquartile ranges are small indicating that the *I/D* values are fairly robust to ±5% error in the flow values. We divided these interquartile ranges by the actual *I/D* values and multiplied by 100 to determine the percent uncertainty in *I/D*. The min is 1.18%, median 2.47%, and max 5.56%. This suggests that the uncertainty in *I/D* is less than the range of uncertainty (10%) in the data used to construct the models. Most importantly, this uncertainty in *I/D* does not lead to a change in the qualitative interpretation of the ratio for any of the models. However, when we consider the min and max values instead of the 1st and 3rd quartiles, we do see that the qualitative interpretation can change for the Cone Springs and Northern Benguela Estuary models from indirect flow dominance to direct flow dominance (Figure 2).

### 3.4 Microbial Food Web Contributions to System Behavior

The microbial food web mediates a large percentage of cycling in many of the models in Model Set 2 (Figure 4). The highest percentage is for the Florida Bay wet season model with 58.2% and the lowest is Swartkops Estuary at 0%. The median value for all models in Figure 4 is 30.2% (Bothnian Bay), and in 17/35 models over a quarter of the cycling is attributed to the microbial food web components. There was no significant linear relationship between IFI and microbial food web FCI ($r^2 = 0.014$, p = 0.50) and results of Spearman's correlation test were non-significant (p = 0.42; ρ = -0.14).

Sites 1 and 2 within the St. Marks seagrass show dominance of indirect flows while sites 3 and 4 do not. To explain this difference, we used EC analysis on these six models (Figure 5) to identify potential differences in the roles of microbial food web compartments. We first observed that where most of the species have a similar role in generating TST, there are a few dominant nodes that contribute to the system behavior. In sites 1 and 2 the top three nodes contributing to total system throughflow are sediment POC, benthic bacteria, and meiofauna, except for in site 2 (Jan) the third position is taken by suspended POC. Benthic bacteria mediate an average of 9.8% of TST in site 1 and 7.8 % in site 2. In site 3 the top three nodes generating system activity are sediment POC, predatory gastropods, and spot; in site 4 these nodes are sediment POC, benthic bacteria, and predatory gastropods. Benthic bacteria only mediates 2.6% of TST in site 3 and 4.5% in site 4. This analysis shows that the impact of the microbial food web is reduced in sites 3 and 4 when compared to sites 1 and 2.

# 4 Discussion

In this paper we tested the generality of the dominance of indirect flows hypothesis on 50 empirically-based ecosystem models using NEA. We found that indirect flows dominate in the majority but not all of the models. When the hypothesis was tested on a subset of the models that have the model structure and species necessary to better represent energy–matter cycling, the evidence is stronger. We also tested the hypotheses relating to the magnitude of indirect flow, and we found no clear relationship between *IFI* and the magnitude of system size and connectance. There is a significant correlation, however, between *IFI* and direct flow intensity and cycling. Our analysis suggests that our NEA results are robust to model uncertainty. Lastly, the results of the node level FCI and EC support the hypothesis that the microbial food web is an important factor in the dominance of indirect effects. The amount of cycling mediated by the microbial food web cannot be used to predict the amount of indirect flow, yet the observed range in this value reflects the variation predicted by the literature (as in Hart et al., 2000). In sum, our evidence supports that indirect energy flows are just as significant, if not more so, than the observed direct feeding behavior, as seen in this study supporting the dominance of indirect flows hypothesis.

## 4.1 Confidence in NEA Results

The qualitative interpretation of the *I/D* ratios can be made with reasonable confidence, as shown by the results of the uncertainty analysis. None of the *I/D* values for any of the models change its qualitative interpretation when the error is introduced using the interquartile range and only two when using the full range. Interestingly, the uncertainty in the NEA results is less than the uncertainty introduced into the data. This is similar to the patterns found previously for network homogenization, another whole system property (Borrett and Salas 2010), and through different methods of analyzing uncertainty (Borrett and Osidele 2007). Further, this is similar to results reported for the Ascendant Perspective of Ecological Network Analysis (ENA) (Kaufman and Borrett, 2010; Kones et al., 2009).

While it would be ideal to test a larger percent error, doing so consistently for all models was prohibited by the thermodynamic constraints of the analysis. Therefore, our conclusions of the robustness of the results of NEA to data uncertainty are limited to a smaller error range. However, we did observe that variation remained small for successful attempts using larger percent errors (data not shown).

### 4.2 Evidence for Dominance of Indirect Flows

The original dominance of indirect effects hypothesis (Higashi and Patten, 1986, 1989; Patten, 1982) suggested that indirect flows should tend to dominate in the majority of ecosystem networks. Our work supports this hypothesis as 74% of all the models had $I > D$. However, all the models in our dataset may not all be suitable to address our specific question. While these models may have been appropriate to address the questions for which they were designed, it does not imply that they can be used broadly for all network questions. Fath (2007) provides guidelines for network construction, one of which is exhaustiveness. A model can only be exhaustive if system components are aggregated together which results in diverse and maybe heterogeneous nodes. However, ecosystem models of energy–matter should explicitly include the species that mediate the important ecosystem processes. Decomposition is one such process, which is critical in the cycling of carbon and ecosystem functioning. The explicit representation of detrital compartments has been shown to influence the results of trophic modeling (e.g. Allesina et al., 2005; Edwards, 2001; Fath and Halnes, 2007). Also vital to the adequate representation of carbon flow are the species that mediate this decomposition and their microbial consumers, and species that consume small phytoplankton. We found that models with some explicit representation of the microbial food web tended to have more recycling and therefore more indirect effects. Figure 5 shows that a significant amount of cycling is indeed being mediated by the microbial food web components in many of the models in Model Set 2. Further, 10 out of the 15 models excluded from Model Set 2 had $I/D < 1$, while only 4/35 had $I/D < 1$ in Model Set 2.

### 4.3 Failures of the Dominance of Indirect Flows Hypothesis in Model Set 2

Despite the tendency for $I/D$ to exceed unity, some of the models in Model Set 2 had direct flows that exceeded indirect flows. To investigate these failures we used the idea that communities that export or bury the products of primary production are typically considered autotrophic, as these products are less available for heterotrophic use (Duarte and Cedbrián, 1996). We would expect autotrophic systems to have a higher tendency to fail the dominance of indirect flows hypothesis as a loss of energy-matter means there is less available for recycling, thus less indirect flow. While productivity estimates were not published for all models and we were not able to use them widely in this work, these ideas are useful in the potential explanation of the failure of the hypothesis.

The St. Marks seagrass, sites 3 and 4, are two of the four models that reveal an $I/D < 1$ in Model Set 2. These, along with the other two St. Marks sites, provide an example of a system where the underlying ecology may legitimately contribute to a lack of indirect flow dominance. A combination of site location and production and cycling measures can help to explain our findings. First, the location of these sites can provide useful information. The sites that exhibit $I>D$ (Figure 3) are located within the relatively protected Goose Creek Bay (Baird et al., 1998). The sites that show $I<D$ (Figure 3) are located in more flow-thru areas; site 3 is located within the mouth of the St. Marks River, and site 4 is located just outside of this mouth (Baird et al., 1998). Supporting this, site 3 has a low residence time compared to the other sites (Baird et al. 1998). Looking at production measures, we see that site 3 has greater production and a larger production/biomass ratio (Baird et al. 1998), indicating that the products of production are likely exiting the system and cannot be recycled at that site. Finally, we also see that the average system level and microbial food web FCI for sites 1 and 2 is greater than for sites 3 and 4 (Table 1 and Figure 5). Thus, while the cycling in this system is generally low, there is clearly more cycling occurring at sites 1 and 2 and a significant portion of this is being mediated by bacterial decomposition.

Taken together, site 3 is a largely autotrophic community where the products of primary production are quickly exported from the system and cannot be utilized by the indirect pathways. Site 4 is located just outside the mouth of the St. Marks River, and while the residence time and measures of production are not notably different than sites 1 and 2, it can be reasoned that its proximity to the river, versus the bay, contribute to its likeness to site 3 versus sites 1 and 2 in regards to cycling and *I/D*. In both sites 3 and 4 the use of recycled carbon is not as prevalent, and therefore the relative role of bacteria in generating TST is not nearly as important. This diminished role of bacteria at sites 3 and 4 is seen in both the FCI decomposition (Figure 4) and the average EC (Figure 5). This congruence further supports the utility of the EC metric (Fann, 2009). In conclusion, the failure of dominance of indirect flow in sites 3 and 4 of the St. Marks seagrass system is likely due to the ecological dynamics at these sites. In addition, the inter-site comparisons reveal the potential role of bacteria in contributing to the magnitude of indirect flows.

The Cypress models (wet and dry) also have $I/D < 1$. In this system detritus is largely buried (Ulanowicz et al., 1997), and thus the use of cycled carbon may not be significant, leading to diminished indirect flows. Therefore, this site can be considered autotrophic and like St. Marks seagrass sites 3 and 4. The representation of the microbial components in this model may also be problematic. The microbial food web is compressed into "living POC", which includes bacteria, microprotozoans, and zooplankton, and "living sediments", which includes bacterioplankton, microfauna, and meiofauna. This aggregation hinders the ability of the flows within the microbial food web to contribute to the sum of indirect flows, and therefore may lead to this sum being smaller than it would be if the microbial food web components were not so highly aggregated. However, this suggests that our analytical results are conservative. This idea is further supported by Fath (2004) who demonstrated that indirect flows tend to increase with system size, though our results and those of Baird et al. (2009) do not match.

This extreme aggregation of the microbial food web occurs in other models in which the $I/D > 1$ (e.g. Graminoids wet and dry) so it may not be the sole cause of $I/D < 1$. This suggests that the manner in which the microbial components are represented may matter, as opposed to just if they are or are not included in the model. The detrital fluxes may also be a problem as flows to detritus are often determined based on the balancing procedure as opposed to empirically determined (e.g. Baird and Milne, 1981; Baird et al., 1991; Miehls et al., 2009ab). Similarly, the dynamics within the microbial food web are difficult to determine empirically (Azam et al., 1983; Baird et al., 2009) and these compartments have not generally been the focus of many studies. In sum, the decomposition process and the role of the microbial food web are to some degree speculation, leading to less confidence in the results presented in Figure 5. Despite the challenges of representing the microbial food web and detritus, the degree of variation in the role of the microbial food web we found is supported by the literature (as in Hart et al., 2000; Sommer et al., 2002) and the results here reinforce the importance of its inclusion in ecosystem models.

### 4.4 Drivers of I/D Variation

An interesting result of NEA on all 50 ecosystem models is the lack of support for two of the four NEA hypotheses regarding under which conditions indirect flow intensity should increase in magnitude. There was no obvious relationship between indirect flow intensity and the size of the system and connectance but there is a significant correlation between IFI and both DFI and FCI. All four variables are predicted by the algebra of NEA to be positively correlated with the magnitude of IFI, and Fath (2004) shows a positive relationship between size of the system and indirect effects using cyber-ecosystems. However, results garnered from the use of ecosystem models do not support these hypotheses. Our results only support the predictions for DFI and FCI.

Our results are similar to previous studies. For example, analysis by Baird et al. (2009) also does not support the predicted relationship between *I/D* and *n*. In addition, in an analysis of the same data set used here Borrett and Salas (2010) did not find evidence of the predicted relationship between homogenization and system size and connectance, and marginal support for the relationship between homogenization and FCI. These results taken together suggest there is potentially an aspect of real ecosystems that is not being captured in the algebraic predictions and or the cyber-ecosystem community assembly model (Fath 2004; Fath and Killian 2007). Fath's original models (2004) are limited in their ability to reflect real ecosystems because they are overly simplified. There are no discernible differences in the species within the defined functional groups, which is not an accurate reflection of natural systems since we would expect some sort of niche differentiation. In addition, the flow values appear to have been randomly assigned in these networks from a uniform random distribution. Despite these simplifications, these models supported Higashi and Patten's (1989) algebraic hypotheses. Again, a key contribution of our work is that we tested the hypotheses on empirically derived ecosystem models.

### 4.5 Methodological Challenges

There are limitations of our model database that could influence our results. These potential problems largely stem from the subjective nature of model construction. For example, our models may be biased due to the limited authorship of the models in our database. The act of constructing an ecosystem model entails many decisions that could affect the results of the analysis (Abarca-Arenas and Ulanowicz, 2002; Allesina et al., 2005; Pinnegar et al., 2005), and this may bias the overall results if a large percentage of the models are constructed by the same author or group of authors. Despite this potential, there is observable variation in the *I/D* values for models constructed by the same author (e.g. Baird et al., 1991; Baird et al., 1998), suggesting that modeling decision bias does not necessarily obscure real ecological differences. A second issue is that the models have different aggregation schemes. Some models such as Lake Findely (Richey et al., 1978) have an extremely high degree of aggregation while others such as Lake Oneida (Miehels et al., 2009) have much finer resolution. These differences in model structure make it impossible to compare results between models (Baird et al., 1991), and might effect the overall conclusions of the study. We doubt, however, that this potential bias would change our overall results. Another limitation is our focus on trophically based models; we excluded more biogeochmically based ecosystem models from this study. Previous results, however, suggest that indirect flows are even larger in biogeochemically based ecosystems models (Borrett et al., 2006; Borrett and Osidele, 2007; Borrett et al. 2010), but this remains to be tested in a systematic fashion. Lastly, where our sample size of 50 models is the largest assembled, it is still a relatively small sample. Despite these limitations, we argue that the model database used to test our hypotheses is a useful data set that is large enough to provide informative results. Despite those issues, we believe that this data is sufficient to address our hypotheses, and these results are supported by our uncertainty analyses.

## 5 Conclusions

In this paper we present evidence that indirect flows tend to dominant direct flows in ecosystems. We also demonstrate the importance of including the microbial food web in ecosystem energy models as this subweb is critical in mediating the flow of carbon in some ecosystems. Lastly, we found that our results were robust to error in the flow estimates, suggesting confidence in our results despite the difficulties of model construction. Our core contribution with this paper is the demonstration of the generality of the dominance of indirect effects in empirically derived and trophically based ecosystem networks. We show that when significant cycling was present in the model nearly every system examined had indirect flows dominating direct flows. These results work in concert with several recent studies to provide strong support for the hypothesis. For example, two studies suggest that relaxing the steady-state assumption of NEA would generate similar results. Borrett et al. (2006) found that indirect flows remain dominant in the Neuse River Estuary despite temporal variability, and Borrett et al. (2010) show that indirect flows develop very rapidly in ecosystems, indicating that an ecosystem does not need to remain at a certain configuration for long before indirect flows will be dominant. This latter conclusion was made using both trophic and biogeochemical networks, indicating that the results are not restricted to trophically-based ecosystem models, as is used in this work. Our findings, in combination with the initial algebra (Higashi and Patten, 1986, 1989), theoretical work (Fath, 2004), and the increasingly common empirical evidence, provides strong support for the systems ecology hypothesis that indirect flows dominate direct flows in ecosystems.

## 6 Acknowledgments

We would like to thank M. Freeze for his help developing the methods for the uncertainty analysis, S.L. Fann for help with environ centrality, and B.C. Patten for review of the manuscript. AKS was supported by the James F. Merritt fellowship from the UNCW Center for the Marine Science and SRB was supported in part by UNCW.

## 7 References

Abarca-Arenas, L.G., Ulanowicz, R.E., 2002. The effects of taxonomic aggregation on network analysis. Ecol. Model. 149, 285–296.

Abrams, P.A., Menge, B.A., Mittlebach, G.G., Spiller, D., Yodzis, P., 1996. The role of indirect effects in food webs. In: Polis G, Winemiller KO, editors. Food Webs: Dynamic and Structure. New York: Chapman and Hall. p371–395.

Azam F., Fenchel T., Field J., Gray J., Meyer-Reid L., and Thingstad F., 1983. The ecological role of water-column microbes in the sea. Mar. Ecol.: Prog. Ser. 10, 257–263.

Allesina, S., and Bondavalli, C., 2003. Steady state of ecosystem flow networks: a comparison between balancing procedures. Ecol. Model. 165, 231–239.

Allesina, S., Bondavalli, C., Scharler, U.M., 2005. The consequences of the aggregation of detritus pools in ecological networks. Ecol. Model. 189, 221–232.

Almunia, J., Basterretxea, G., Aistegui, J., Ulanowicz, R.E., 1999. Benthic-pelagic switching in a coastal subtropical lagoon. Estuarine, Coastal Shelf Sci. 49, 363–384.

Baird, D., Asmus, H., Asmus, R., 2004a. Energy flow of a boreal intertidal ecosystem, the Sylt-Rømø Bight. Mar. Ecol.: Prog. Ser. 279, 45−61.

Baird, D., Asmus, H., Asmus, R., 2008. Nutrient dynamics in the Sylt-Romo Bight ecosystem, German Wadden Sea: an ecological network analysis approach. Estuarine, Coastal Shelf Sci. 80, 339−356.

Baird, D., Christian, R.R., Peterson, C.H., Johnson, G.A., 2004b. Consequences of hypoxia on estuarine ecosystem function: energy diversion from consumers to microbes. Ecol. Appl. 14, 805−822.

Baird, D., Fath, B.D., Ulanowicz, R.E., Asmus, H., Asmus, R., 2009. On the consequences of aggregation and balancing of networks on system properties derived from ecological network analysis. Ecol. Model. 220, 3465−3471.

Baird, D., Luczkovich, J., Christian, R.R., 1998. Assessment of spatial and temporal variability in ecosystem attributes of the St. Marks National Wildlife Refuge, Apalachee Bay, Florida. Estuarine, Coastal Shelf Sci. 47, 329−349.

Baird, D., McGlade, J.M., and Ulanowicz, R.E., 1991. The comparative ecology of six marine ecosystems. Philos. T. Roy. Soc. B. 333, 15−29.

Baird, D., and Milne, H., 1981. Energy flow in the Ythan Estuary, Aberdeenshsire, Scotland. Estuarine, Coastal Shelf Sci. 13, 455−472.

Baird, D., Ulanowicz, R.E., 1989. The seasonal dynamics of the Chesapeake Bay ecosystem. Ecol. Monogr. 59, 329−364.

Bever, J.D., 1999. Dynamics within mutualism and the maintenance of diversity: inference from a model of interguild frequency dependence. Ecol. Lett. 2, 52−62.

Bever, J.D., 2002. Negative feedback within a mutualism: host-specific growth of mycorrhiza fungi reduced plant benefit. P. Roy. Soc. Lond. B. Bio. 269, 2595−2601.

Borgatti, S.P., Everett, M.G., 2006. A graph-theoretic perspective on centrality. Soc. Networks 28, 466−484.

Bondavalli, C., Ulanowicz, R.E., 1999. Unexpected effects of predators upon their prey: the case of the American alligator. Ecosystems 2, 49−63.

Borrett, S.R., Fath, B.D., Patten, B.C., 2007. Functional integration of ecological networks through pathway proliferation. J. Theor. Biol. 245, 98−111.

Borrett, S.R., Salas, A.K., 2010. Evidence for resource homogenization in 50 trophic ecosystem networks. Ecol. Model. 221, 1710−1716

Borrett, S.R., Osidele, O.O., 2007. Environ indicator sensitivity to flux uncertainty in a phosphorus model of Lake Sidney Lanier, USA. Ecol. Model. 200, 371−383.

Borrett, S.R., Whipple, S.J., Patten, B.C., Christian, R.R., 2006. Indirect effects and distributed control in ecosystems: temporal variation of indirect effects in a seven-compartment model of nitrogen flow in the Neuse River Estuary, USA−time series analysis. Ecol. Model. 194, 178−188.

Borrett, S.R., Whipple, S. J., Patten, B.C., 2010. Rapid development of indirect effects in ecological networks. Oikos 119, 1136−1148.

Brylinsky, M., 1972. Steady-state sensitivity analysis of energy flow in a marine ecosystem. In B.C. Patten, editor. Systems Analysis and Simulation in Ecology Volume 2. Academic Press, New York. p81−101.

Christian, R.R., Forés, E., Comin, F., Viaroli, P., Naldi, M., Ferrari, I., 1996. Nitrogen cycling networks of coastal ecosystems: influence of trophic status and primary producer form. Ecol. Model. 87, 111−129.

Dame, R.F., Patten, B.C., 1981. Analysis of energy flows in an intertidal oyster reef. Mar. Ecol.: Prog. Ser. 5, 115−124.

Diekötter, T., Haynes, K.J., Mazeffa, D., Crist, T.O., 2007. Direct and indirect effects of habitat area and matrix composition on species interactions among flower-visiting insects. Oikos 116, 1588−1598.

Duarte, C. M., Cebrián, J., 1996. The fate of marine autotrophic production. Limnol. Oceanogr. 41, 1758−1766.

Edwards, A.M., 2001. Adding detritus to a nutrient-phytoplankton-zooplankton model: a dynamical-systems approach. J. Plankton Res. 23, 389−413.

Estrada, E., 2010. Generalized walks-based centrality measures for complex biological networks. J. Theor. Biol. 263, 556−565.

Fann, S.F., 2009. Environ centrality quantifies the relative roles of species in generating ecosystem activity. Honors Thesis, University of North Carolina Wilmington.

Fath, B.D., 2004. Network analysis applied to large-scale cyber-ecosystems. Ecol. Model. 171, 329−337.

Fath, B.D., Borrett, S.R., 2006. A MATLAB® function for Network Environ Analysis. Environ. Modell. Softw. 21, 375−405.

Fath, B.D., Halnes, G., 2007. Cyclic energy pathways in ecological food webs. Ecol. Model. 208, 17−24.

Fath, B.D., Killian, M.C., 2007. The relevance of ecological pyramids in community assemblages. Ecol. Model. 208, 286−294.

Fath, B.D., Patten, B.C., 1998. Network synergism: emergence of positive relations in ecological systems. Ecol. Model. 107, 127−143.

Fath, B.D., Patten, B.C., 1991. Review of the foundations of network environ analysis. Ecosystems 2, 167−179.

Fath, B.D., Scharler, U.M., Ulanowicz, R.E., Hannon, B., 2007. Ecological network analysis: network construction. Ecol. Model. 208, 49−55.

Fenchel, T., 1970. Studies on the decomposition of organic detritus derived from the turtle grass *Thalassia testudinum*. Limnol. Oceanogr. 15, 14−20.

Finn, J.T., 1976. Measures of ecosystem structure and function derived from analysis of flows. J. Theor. Biol. 56, 363−380.

Finn, J.T., 1980. Flow analysis of models of the Hubbard brook ecosystem. Ecology 61, 562−571.

Hanneman, R.A., Riddle, M.A. Centrality and Power *in* Introduction to social network methods. (Riverside, CA: University of California, Riverside, 2005), http://faculty.ucr.edu/~hanneman/ (accessed October 10, 2009).

Hart, D.R., Stone, L., Berman, T., 2000. Seasonal dynamics of the Lake Kinneret food web: The importance of the microbial loop. Limnol. Oceanogr. 45, 350−361.

Heymans, J.J., Baird, D., 2000. A carbon flow model and network analysis of the Northern Benguela Upwelling system, Namibia. Ecol. Model. 126, 9−32.

Higashi, M., Patten, B.C., 1986. Further aspects of the analysis of indirect effects in ecosystems. Ecol. Model. 31, 69−77.

Higashi, M., Patten, B.C., 1989. Dominance of indirect causality in ecosystems. Am. Nat. 133, 288–302.

Jørgensen, S.E., Fath, B.D., Bastianoni, S., Marques, J., Muller, F., Nielsen, S.N., Patten, V., Tiezzi, E., Ulanowicz, R., 2007. A New Ecology: Systems Perspective. Netherlands: Elsevier. 288p.

Jørgensen, S.E., Patten, B.C., Straskraba, M., 1992. Ecosystems emerging - toward an ecology of complex-systems in a complex future. Ecol. Model. 62, 1–27.

Jørgensen, S.E., Patten, B.C., Straskraba, M., 1999. Ecosystems emerging: 3. Openness. Ecol. Model. 114, 41–64.

Kaufman, A.G., Borrett, S.R., 2010. Ecosystem network analysis indicators are generally robust to parameter uncertainty in a phosphorus model of Lake Sidney Lanier, USA. Ecol. Model. 221, 1230–1238

Kones, J.K., Soetaert, K., van Oevelen, D., Owino, J.O., 2009. Are network indices robust indicators of food web functioning? A Monte Carlo approach. Ecol. Model. 220, 370–382.

Larsson, U., Hagström, Å., 1982. Fractionated phytoplankton primary production, exudates release, and bacterial production in a Baltic eutrophication gradient. Mar. Biol. 67, 57–70.

Link, J., Overholtz, W., O'Reilly, J., Green, J., Dow, D., Palka, D., Legault, C., Vitaliano, J., Guida, V., Fogarty, M., Brodziak, J., Methratta, L., Stockhausen, W., Col, L., Griswol, C., 2008. The northeast U.S. continental shelf energy modeling and analysis exercise (EMAX): Ecological network model development and basic ecosystem metrics. J. Marine Syst. 74, 453–474.

Mann, K.H., 1988. Production and use of detritus in various freshwater, estuarine, and coastal marine ecosytems. Limnol. Oceanogr. 33, 910–930.

Menge, B.A., 1997. Detection of direct versus indirect effects: were experiments long enough? Am. Nat.149, 801–823.

Miehls, A.L.J., Mason, D.M., Frank, K.A., Krause, A.E., Peacor, S.D., Taylor, W.W., 2009a. Invasive species impacts on ecosystem structure and function: A comparison of Oneida Lake, New York, USA, before and after zebra mussel invasion. Ecol. Model. 220, 3194–3209.

Miehls, A.L.J., Mason, D.M., Frank, K.A., Krause, A.E., Peacor, S.D., Taylor, W.W., 2009b. Invasive species impacts on ecosystem structure and function: A comparison of the Bay of Quinte, Canada and Oneida Lake, USA, before and after zebra mussel invasion. Ecol. Model. 220, 3182–3193.

Monaco, M.E., Ulanowicz, R.E., 1997. Comparative ecosystem trophic structure of three U.S. Mid-Atlantic estuaries. Mar. Ecol.: Prog. Ser. 161, 239–254.

Odum, H.T., 1957. Trophic structure and productivity of Silver Springs, Florida. Ecol. Monogr. 27, 55–112.

Patten, B.C., 1982. Environs: Relativistic elementary particles for ecology. Am. Nat. 119, 179–219.

Patten, B.C., 1983. On the quantitative dominance of indirect effects in ecosystems. In W.K. Lauenrtoh, G.V. Skogerboe, and M. Flug, editors. Analysis of Ecological Systems: State-of-the-Art in Ecological Modelling. Elsevier, Amsterdam. p27–37.

Patten, B.C., 1984. Further developments toward a theory of the quantitative importance of indirect effects in ecosystem. Verh. Gesellschaft für Ökologie 13, 271–284.

Patten, B.C., 1985. Energy cycling in the ecosystem. Ecol. Model. 28, 1–71.

Patten B.C., Bosserman, R.W., Finn, J.T., Cale, W.G., 1976. Propagation of cause in ecosystems. In Patten, B.C., editor. Systems Analysis and Simulation in Ecology, Volume IV. Academic Press, New York. p457−579.

Patten, B.C., Higashi, M., Burns, T., 1990. Trophic dynamics in ecosystem networks: significance of cycles and storage. Ecol. Model. 51, 1−28.

Patten, B.C., 1991. Network ecology: indirect determination of the life-environment relationship in ecosystems. Higashi, M., Burns, T.P., editors. Theoretical Studies of Ecosystems: The Network Perspective. Cambride: Cambridge University Press. p289−351

Pinnegar, J.K., Blanchard, J.L., Mackinson, S., Scott, R.D., Duplisea, D.E., 2005. Aggregation and removal of weak-links in food web models: system stability and recovery from disturbance. Ecol. Model. 184, 229−248.

Pomeroy, L.R., 1974. The ocean's food web, a changing paradigm. BioScience 24, 499−504.

Preisser, E.L., Bolnick, D.I., Benard, M.F., 2005. Scared to death? The effects of intimidation and consumption in predator-prey interactions. Ecology 86, 501−509.

Richey, J.E., Richey, J.E., Wissmar, R.C., Devol, A.H., Likens, G.E., Eaton, J.S., Wetzel, R.G., Odum, W.E., Johnson, N.M., Loucks, O.L., Prentki, R.T., Rich, P.H., 1978. Carbon flow in four lake ecosystems: a structural approach. Science 202, 1183−1186.

Rybarczyk, H., Elkaim, B., Ochs, L., Loquet, N., 2003. Analysis of the trophic network of a macrotidal ecosystem: the Bay of Somme (Eastern Channel). Estuarine, Coastal Shelf Sci. 58, 405−421.

Sandberg, J., Elmgren, R., Wulff, F., 2000. Carbon flows in Baltic sea food webs-a re-evaluation using a mass balance approach. J. Marine. Syst. 25, 249−260.

Sherr, E., Sherr, B., 1988. Role of microbes in pelagic food webs: a revised concept. Limnol. Oceanogr. 33, 1225−1227.

Sommer, U., Stibor, H., Katechakis, A., Sommer, F., Hansen, T., 2002. Pelagic food web configurations at different levels of nutrient richness and their implications for the ratio fish production: primary production. Hydrobiologia 484, 11−20.

Strauss, S.Y., 1991. Indirect effects in community ecology: their definition, study and importance. Trends Ecol. Evol. 6, 206−210.

Tilly, L.J., 1968. The structure and dynamics of Cone Spring. Ecol. Monogr. 38, 169−197.

Turner, A., Mittelbach, G., 1990. Predator avoidance and community structure: interactions among piscivores, planktivores, and plankton. Ecology 71, 2241−2254.

Ulanowicz, R.E., 1986. Growth and Development: Ecosystems Phenomenology. toExcel Press, Lincoln.

Ulanowciz, R.E., 1995. Trophic flow networks as indicators of ecosystem stress. In Polis, G.A. and Winemiller, K.O., editors. Food Webs: Integration of Patterns and Dynamics. Chapman and Hall, NY. p358−368.

Ulanowicz, R.E., Bondavalli, C., Egnotovich, M.S., 1997. Network analysis of trophic dynamics in South Florida ecosystem, FY 96: The Cypress wetland ecosystem. Annual Report to the United States Geological Service Biological Resources Division Ref. No. [UMCES]CBL 97-075, Chesapeake Biological Laboratory, University of Maryland.

Ulanowicz, R.E., Bondavalli, C., Egnotovich, M.S., 1998. Network analysis of trophic dynamics in South Florida ecosystem, FY 97: The Florida Bay ecosystem. Annual Report to the United States Geological Service Biological Resources Division Ref. No. [UMCES] CBL 98-123, Chesapeake Biological Laboratory, University of Maryland.

Ulanowicz, R.E., Bondavalli, C., Heymans, J.J., Egnotovich, M.S., 1999. Network analysis of trophic dynamics in South Florida ecosystem, FY 98: The mangrove ecosystem. Annual Report to the United States Geological Service Biological Resources Division Ref. No. [UMCES] CBL 99-0073. Technical Report Series No. TS-191-99, Chesapeake Biological Laboratory, University of Maryland.

Ulanowicz, R.E., Bondavalli, C., Heymans, J.J., Egnotovich, M.S., 2000. Network analysis of trophic dynamics in South Florida ecosystem, FY 99: The graminoid ecosystem. Annual Report to the United States Geological Service Biological Resources Division Ref. No. [UMCES] CBL 00-0176, Chesapeake Biological Laboratory, University of Maryland.

Ulanowicz, R.E., Puccia, C.J., 1990. Mixed tropic impacts in ecosystems. Coenoses 5:7−16.

Wasserman, S., Faust, K., 1994. Social Network Analysis: Methods and Applications. Cambridge University Press, Cambridge, U.K. 857p.

Wilkinson, D.M., 2006. Fundamental Processes in Ecology an Earth Systems Approach. Oxford University Press, New York. 200p.

Wootton, J.T., 1993. Indirect effects and habitat use in an intertidal community: interaction chains and interaction modifications. Am. Nat. 141, 71−89.

Wootton, J.T., 1994. The nature and consequences of indirect effects in ecological communities. Annu. Rev. Ecol. Syst. 25, 443−466.

Yodzis, P., 1988. The indeterminacy of ecological interactions as perceived through perturbation experiments. Ecology 69, 508−515.

Table 1: 50 empirically derived and trophically-based ecosystem models. The complete list is Model Set 1 and models excluded from Model Set 2 are indicated with superscript 1 (no cycling) and/or superscript 2 (no microbial food web representation).

| Model | Units | $n^a$ | $C^a$ | $TST^a$ | $FCI^a$ | $I/D^a$ | Source |
|---|---|---|---|---|---|---|---|
| Lake Findley[2] | gC m$^{-2}$ yr$^{-1}$ | 4 | 0.38 | 51 | 0.30 | 1.72 | Richey et al. 1978 |
| Mirror Lake[2] | gC m$^{-2}$ yr$^{-1}$ | 5 | 0.36 | 218 | 0.32 | 1.86 | Richey et al. 1978 |
| Lake Wingra[2] | gC m$^{-2}$ yr$^{-1}$ | 5 | 0.40 | 1517 | 0.40 | 1.86 | Richey et al. 1978 |
| Marion Lake[2] | gC m$^{-2}$ yr$^{-1}$ | 5 | 0.36 | 243 | 0.31 | 2.18 | Richey et al. 1978 |
| Cone Springs[2] | kcal m$^{-2}$ yr$^{-1}$ | 5 | 0.32 | 30626 | 0.09 | 1.02 | Tilly 1968 |
| Silver Springs[1,2] | kcal m$^{-2}$ yr$^{-1}$ | 5 | 0.28 | 29175 | 0.00 | 0.20 | Odum 1957 |
| English Channel[1,2] | kcal m$^{-2}$ yr$^{-1}$ | 6 | 0.25 | 2280 | 0.00 | 0.44 | Brylinsky 1972 |
| Oyster Reef | Kcal m$^{-2}$ yr$^{-1}$ | 6 | 0.33 | 84 | 0.11 | 1.58 | Dame and Patten 1981 |
| Bay of Somme[2] | mgC m$^{-2}$ d$^{-1}$ | 9 | 0.30 | 2035 | 0.14 | 0.74 | Rybarczyk et al. 2003 |
| Bothnian Bay | gC m$^{-2}$ yr$^{-1}$ | 12 | 0.22 | 130 | 0.18 | 1.97 | Sandberg et al. 2000 |
| Bothnian Sea | gC m$^{-2}$ yr$^{-1}$ | 12 | 0.24 | 458 | 0.27 | 2.52 | Sandberg et al. 2000 |
| Ythan Estuary[2] | gC m$^{-2}$ yr$^{-1}$ | 13 | 0.23 | 4181 | 0.24 | 1.89 | Baird and Milne 1981 |
| Baltic Sea | mgC m$^{-2}$ d$^{-1}$ | 15 | 0.17 | 1974 | 0.13 | 1.93 | Baird et al. 1991 |
| Ems Estuary | mgC m$^{-2}$ d$^{-1}$ | 15 | 0.19 | 1019 | 0.32 | 2.95 | Baird et al. 1991 |
| Swarkops Estuary | mgC m$^{-2}$ d$^{-1}$ | 15 | 0.17 | 13996 | 0.47 | 2.73 | Baird et al. 1991 |
| Southern Benguela Upwelling | mgC m$^{-2}$ d$^{-1}$ | 16 | 0.23 | 1774 | 0.19 | 1.24 | Baird et al. 1991 |
| Peruvian Upwelling | mgC m$^{-2}$ d$^{-1}$ | 16 | 0.22 | 33496 | 0.04 | 1.16 | Baird et al. 1991 |
| Crystal River (control)[2] | mgC m$^{-2}$ d$^{-1}$ | 21 | 0.19 | 15063 | 0.07 | 0.67 | Ulanowicz 1986, 1995 |
| Crystal River (thermal)[2] | mgC m$^{-2}$ d$^{-1}$ | 21 | 0.14 | 12032 | 0.09 | 0.71 | Ulanowicz 1986, 1995 |
| Charca de Maspalomas Lagoon | mgC m$^{-2}$ d$^{-1}$ | 21 | 0.13 | 6010331 | 0.18 | 2.69 | Alumunia et al. 1999 |
| Northern Benguela Upwelling | mgC m$^{-2}$ d$^{-1}$ | 24 | 0.21 | 6608 | 0.05 | 1.04 | Heymans and Baird 2000 |
| Neuse Estuary (early summer 1997) | mgC m$^{-2}$ d$^{-1}$ | 30 | 0.09 | 13826 | 0.12 | 1.48 | Baird et al. 2004b |
| Neuse Estuary (late summer 1997) | mgC m$^{-2}$ d$^{-1}$ | 30 | 0.11 | 13038 | 0.13 | 1.31 | Baird et al. 2004b |
| Neuse Estuary (early summer 1998) | mgC m$^{-2}$ d$^{-1}$ | 30 | 0.09 | 14025 | 0.12 | 1.41 | Baird et al. 2004b |
| Neuse Estuary (late summer 1998) | mgC m$^{-2}$ d$^{-1}$ | 30 | 0.10 | 15031 | 0.11 | 1.26 | Baird et al. 2004b |
| Gulf of Maine | g ww m$^{-2}$ yr$^{-1}$ | 31 | 0.35 | 18382 | 0.15 | 2.11 | Link et al. 2008 |
| Georges Bank | g ww m$^{-2}$ yr$^{-1}$ | 31 | 0.35 | 16890 | 0.18 | 2.58 | Link et al. 2008 |
| Middle Atlantic Bight | g ww m$^{-2}$ yr$^{-1}$ | 32 | 0.37 | 17917 | 0.18 | 2.78 | Link et al. 2008 |
| Narragansett Bay | mgC m$^{-2}$ yr$^{-1}$ | 32 | 0.15 | 3917246 | 0.51 | 6.93 | Monaco and Ulanowicz 1997 |
| Southern New England Bight | g ww m$^{-2}$ yr$^{-1}$ | 33 | 0.03 | 17597 | 0.16 | 2.56 | Link et al. 2008 |
| Chesapeake Bay | mgC m$^{-2}$ yr$^{-1}$ | 36 | 0.09 | 3227453 | 0.19 | 2.76 | Baird and Ulanowicz 1989 |
| St. Marks Seagrass, site 1 (Jan) | mgC m$^{-2}$ d$^{-1}$ | 51 | 0.08 | 1316 | 0.13 | 1.70 | Baird et al. 1998 |
| St. Marks Seagrass, site 1 (Feb) | mgC m$^{-2}$ d$^{-1}$ | 51 | 0.08 | 1591 | 0.11 | 1.66 | Baird et al. 1998 |
| St. Marks Seagrass, site 2 (Jan) | mgC m$^{-2}$ d$^{-1}$ | 51 | 0.07 | 1383 | 0.09 | 1.24 | Baird et al. 1998 |
| St. Marks Seagrass, site 2 (Feb) | mgC m$^{-2}$ d$^{-1}$ | 51 | 0.08 | 1921 | 0.08 | 1.30 | Baird et al. 1998 |
| St. Marks Seagrass, site 3 (Jan) | mgC m$^{-2}$ d$^{-1}$ | 51 | 0.05 | 12651 | 0.01 | 0.12 | Baird et al. 1998 |
| St. Marks Seagrass, site 4 (Feb) | mgC m$^{-2}$ d$^{-1}$ | 51 | 0.08 | 2865 | 0.04 | 0.71 | Baird et al. 1998 |
| Sylt Rømø Bight | mgC m$^{-2}$ d$^{-1}$ | 59 | 0.08 | 1353406 | 0.09 | 1.54 | Baird et al. 2004a |
| Graminoids (wet) | gC m$^{-2}$ yr$^{-1}$ | 66 | 0.18 | 13677 | 0.02 | 1.20 | Ulanowicz et al. 2000 |
| Graminoids (dry) | gC m$^{-2}$ yr$^{-1}$ | 66 | 0.18 | 7520 | 0.04 | 1.41 | Ulanowicz et al. 2000 |
| Cypress (wet) | gC m$^{-2}$ yr$^{-1}$ | 68 | 0.12 | 2572 | 0.04 | 0.62 | Ulanowicz et al. 1997 |
| Cypress (dry) | gC m$^{-2}$ yr$^{-1}$ | 68 | 0.12 | 1918 | 0.04 | 0.70 | Ulanowicz et al. 1997 |
| Lake Oneida (pre-ZM)[2] | gC m$^{-2}$ yr$^{-1}$ | 74 | 0.22 | 1638 | <0.01 | 0.17 | Miehls et al. 2009a |
| Lake Quinte (pre-ZM)[2] | gC m$^{-2}$ yr$^{-1}$ | 74 | 0.21 | 1467 | <0.01 | 0.15 | Miehls et al. 2009b |
| Lake Oneida (post-ZM)[2] | gC m$^{-2}$ yr$^{-1}$ | 76 | 0.22 | 1365 | <0.01 | 0.41 | Miehls et al. 2009a |
| Lake Quinte (post-ZM)[2] | gC m$^{-2}$ yr$^{-1}$ | 80 | 0.21 | 1925 | 0.01 | 0.30 | Miehls et al. 2009b |
| Mangroves (wet) | gC m$^{-2}$ yr$^{-1}$ | 94 | 0.15 | 3272 | 0.10 | 1.69 | Ulanowicz et al. 1999 |
| Mangroves (dry) | gC m$^{-2}$ yr$^{-1}$ | 94 | 0.15 | 3266 | 0.10 | 1.70 | Ulanowicz et al. 1999 |
| Florida Bay (wet) | mgC m$^{-2}$ yr$^{-1}$ | 125 | 0.12 | 2721 | 0.14 | 1.73 | Ulanowicz et al. 1998 |
| Florida Bay (dry) | mgC m$^{-2}$ yr$^{-1}$ | 125 | 0.13 | 1779 | 0.08 | 1.29 | Ulanowicz et al. 1998 |

[a] $n$ is the number of nodes in the network model, $C = L/n^2$ is the model connectance when $L$ is the number of direct links or energy-matter transfers, $TST = \sum\sum f_i j + \sum z_i$ is the total system throughflow, $FCI$ is the Finn Cycling Index, and $I/D$ is the indirect/direct flow ratio.

## 8 Figure Legends

**Figure 1**: Hypothetical example ecosystem network with $n = 4$. A) Information regarding who is connected to whom in the network is in the adjacency matrix, **A**. The actual amount of energy flowing between the nodes is found in the flow matrix, **F**. Energy being transported across the system boundary into the nodes is represented in $\vec{z}$, and energy exiting the system across the boundary is in $\vec{y}$. B) Direct flow intensity is represented in **G**, and is the percentage of energy each node receives directly from each other node. When **G** is raised to a power *m*, the resulting matrices represent the indirect flow intensity between the nodes over all the paths of length *m*.

**Figure 2**: Indirect/direct (*I/D*) flow ratio for 50 empirically derived and trophically-based ecosystem models. Black bars are networks composing Model Set 2, those models that satisfy the second set of criteria of cycling structure and representation of the microbial food web. Extension of bars beyond $I/D = 1$ indicates indirect flow dominance. The range of uncertainty between maximum and minimum values generated from our Monte-Carlo type uncertainty analysis is indicated for each model by the error bars. A few of these values have been previously published using an alternate calculation: [1]Jørgensen and Fath 2007; [3]Baird et al. 2007; [4]Borrett et al (2010). Only one value has been published using the calculation we use here: [2]Borrett et al. (2006).

**Figure 3**: Test of the relationship between indirect flow intensity (IFI) and (a) system size, *n*, (b) connectance, *C*, (c) direct flow intensity, DFI, and (d) Finn Cycling Index, FCI. There is no significant linear relationship between IFI and system size ($r^2 = 0.069$; p = 0.066) and connectance ($r^2 = 0.022$; p = 0.31).

**Figure 4**: Rank order percentage of Finn Cycling Index (FCI) that is mediated by components of the microbial food web in the ecosystem models in Model Set 2.

**Figure 5**: Average Environ Centrality for St. Marks seagrass system, site 1 (a, b), site 2 (c, d), site 3 (e), and site 4 (f). Average Environ Centrality is the percentage of TST mediated by each node. The nodes are ranked on the x-axis from highest to lowest Average Environ Centrality, with the three highest ranking nodes indicated.

Figure 1

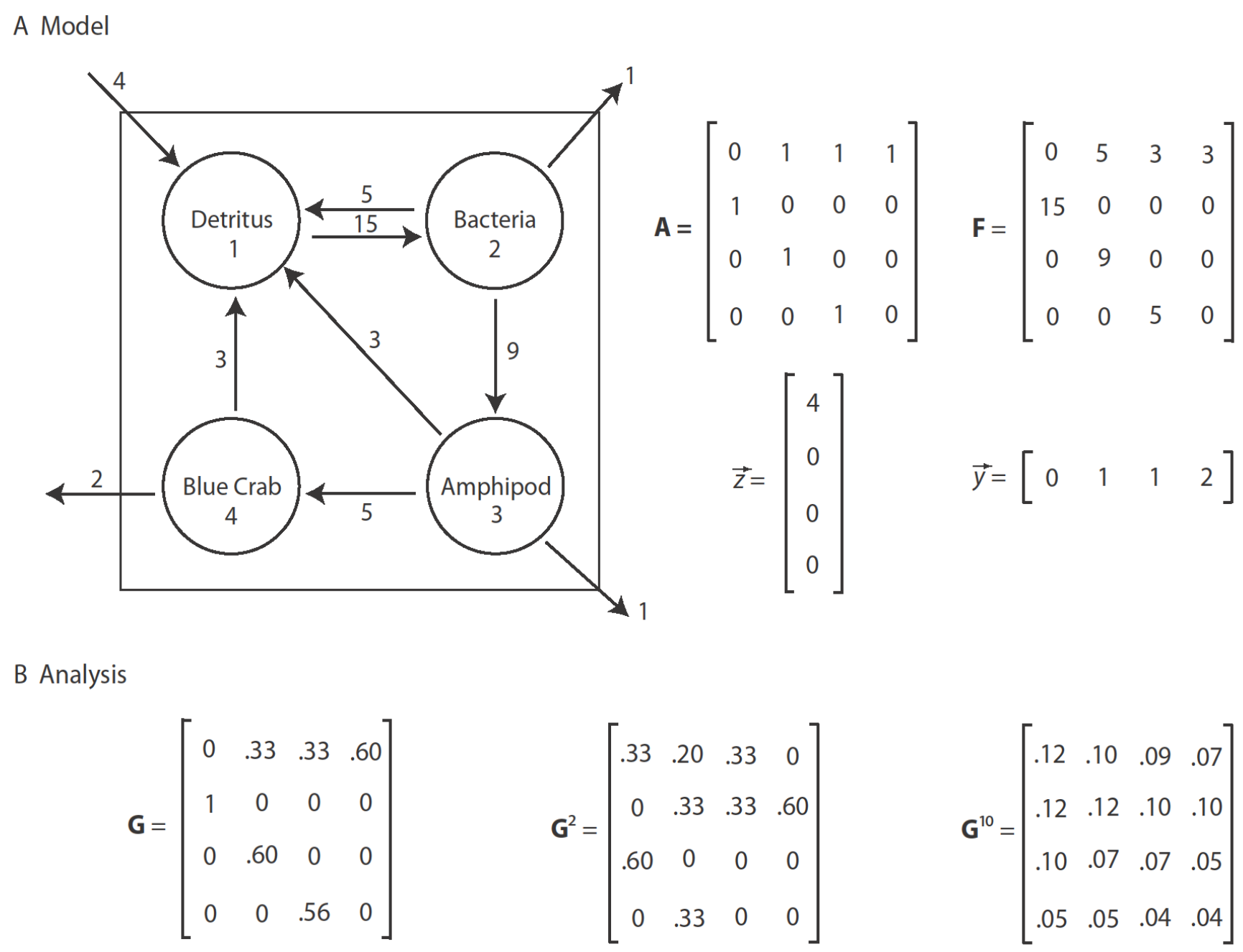
A Model
Detritus 1
Bacteria 2
Amphipod 3
Blue Crab 4
4
1
5
15
3
3
9
2
5
1
A =
0 1 1 1
1 0 0 0
0 1 0 0
0 0 1 0
F =
0 5 3 3
15 0 0 0
0 9 0 0
0 0 5 0
z =
4
0
0
0
y = 0 1 1 2
B Analysis
G =
0 .33 .33 .60
1 0 0 0
0 .60 0 0
0 0 .56 0
G² =
.33 .20 .33 0
0 .33 .33 .60
.60 0 0 0
0 .33 0 0
G¹⁰ =
.12 .10 .09 .07
.12 .12 .10 .10
.10 .07 .07 .05
.05 .05 .04 .04

Figure 2

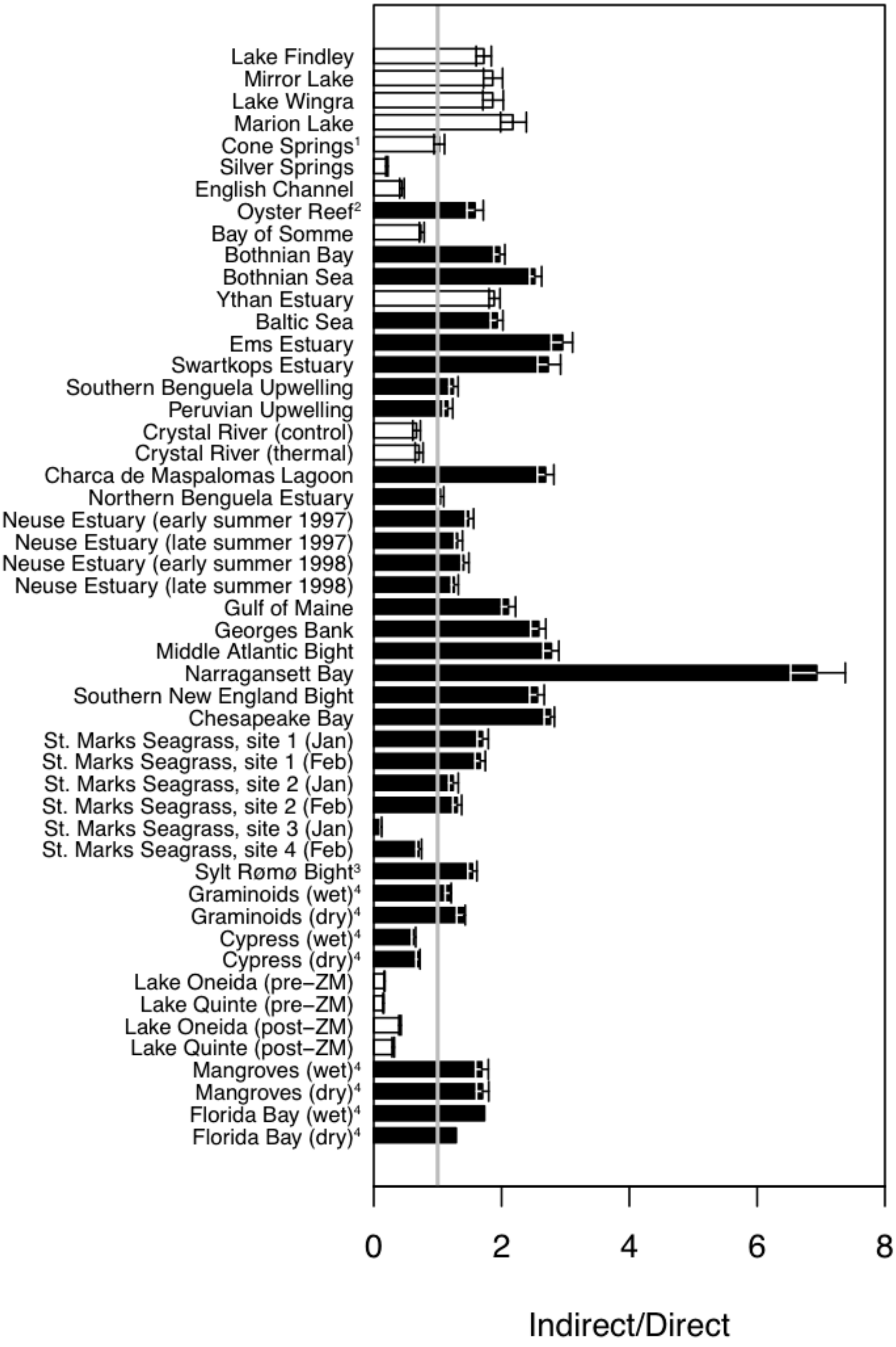
Lake Findley
Mirror Lake
Lake Wingra
Marion Lake
Cone Springs[1]
Silver Springs
English Channel
Oyster Reef[2]
Bay of Somme
Bothnian Bay
Bothnian Sea
Ythan Estuary
Baltic Sea
Ems Estuary
Swartkops Estuary
Southern Benguela Upwelling
Peruvian Upwelling
Crystal River (control)
Crystal River (thermal)
Charca de Maspalomas Lagoon
Northern Benguela Estuary
Neuse Estuary (early summer 1997)
Neuse Estuary (late summer 1997)
Neuse Estuary (early summer 1998)
Neuse Estuary (late summer 1998)
Gulf of Maine
Georges Bank
Middle Atlantic Bight
Narragansett Bay
Southern New England Bight
Chesapeake Bay
St. Marks Seagrass, site 1 (Jan)
St. Marks Seagrass, site 1 (Feb)
St. Marks Seagrass, site 2 (Jan)
St. Marks Seagrass, site 2 (Feb)
St. Marks Seagrass, site 3 (Jan)
St. Marks Seagrass, site 4 (Feb)
Sylt Rømø Bight[3]
Graminoids (wet)[4]
Graminoids (dry)[4]
Cypress (wet)[4]
Cypress (dry)[4]
Lake Oneida (pre−ZM)
Lake Quinte (pre−ZM)
Lake Oneida (post−ZM)
Lake Quinte (post−ZM)
Mangroves (wet)[4]
Mangroves (dry)[4]
Florida Bay (wet)[4]
Florida Bay (dry)[4]
0
2
4
6
8
Indirect/Direct

Figure 3

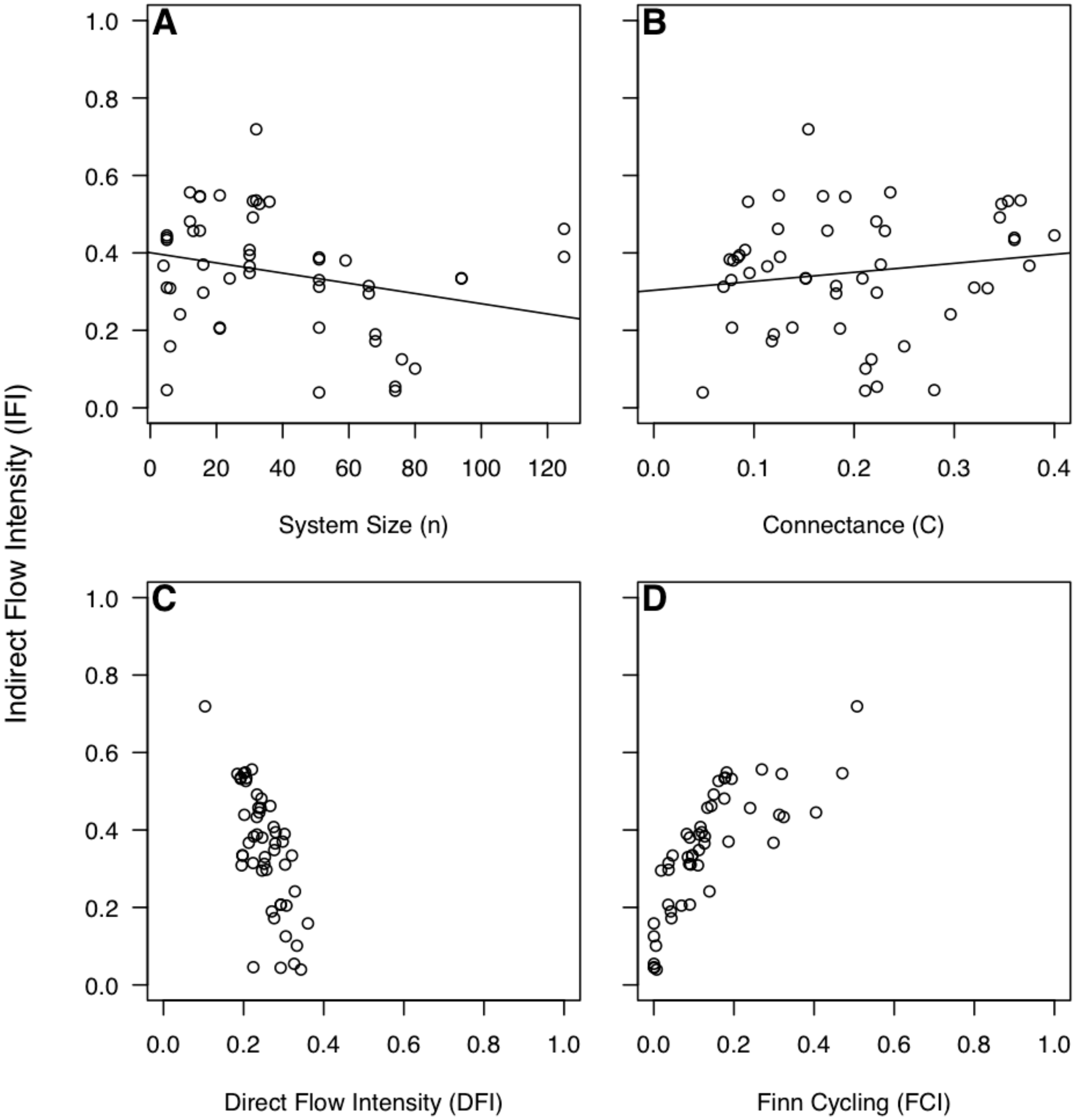
A
B
C
D
Indirect Flow Intensity (IFI)
System Size (n)
Connectance (C)
Direct Flow Intensity (DFI)
Finn Cycling (FCI)

837 Figure 4

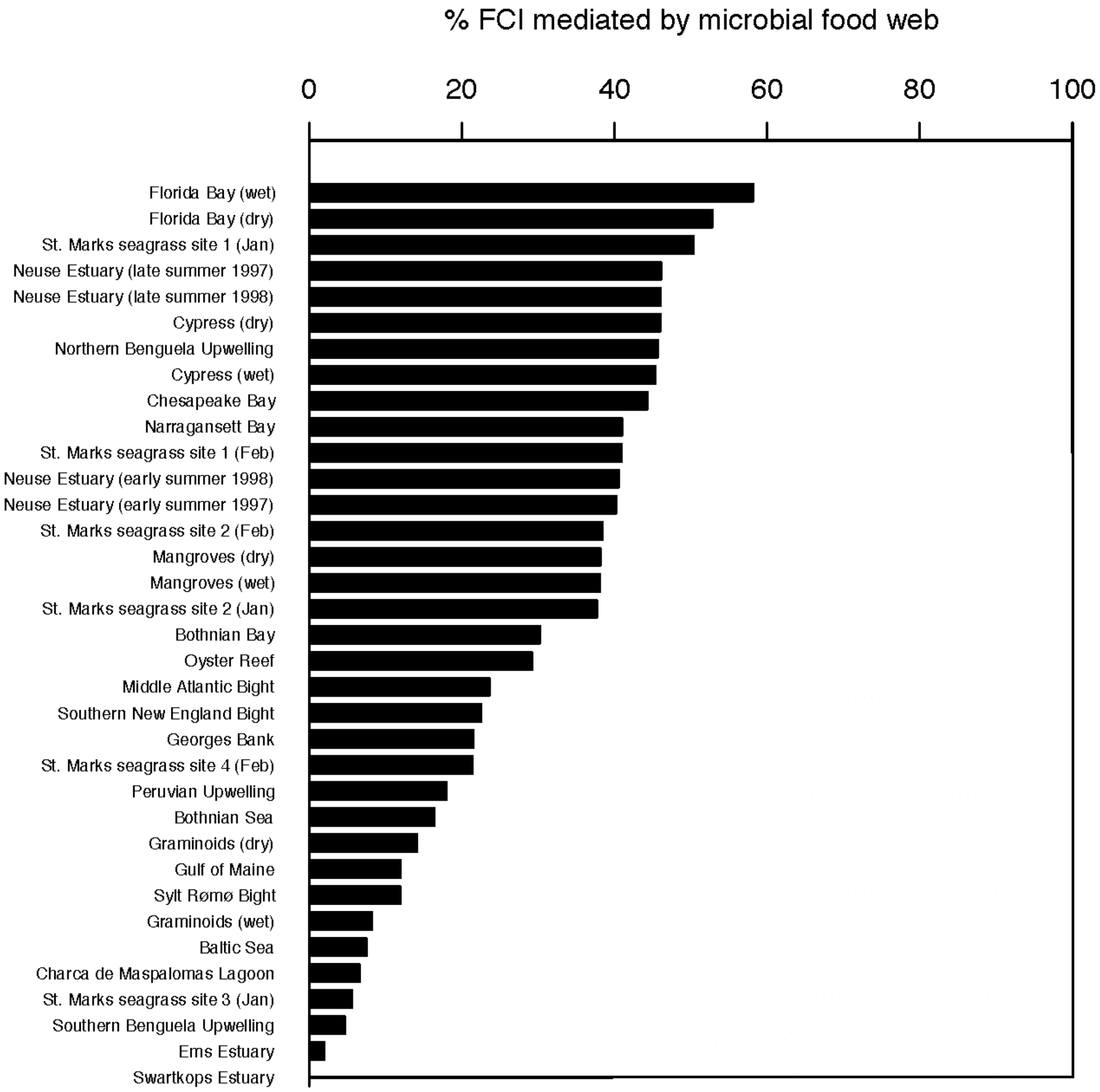

838

839 Figure 5

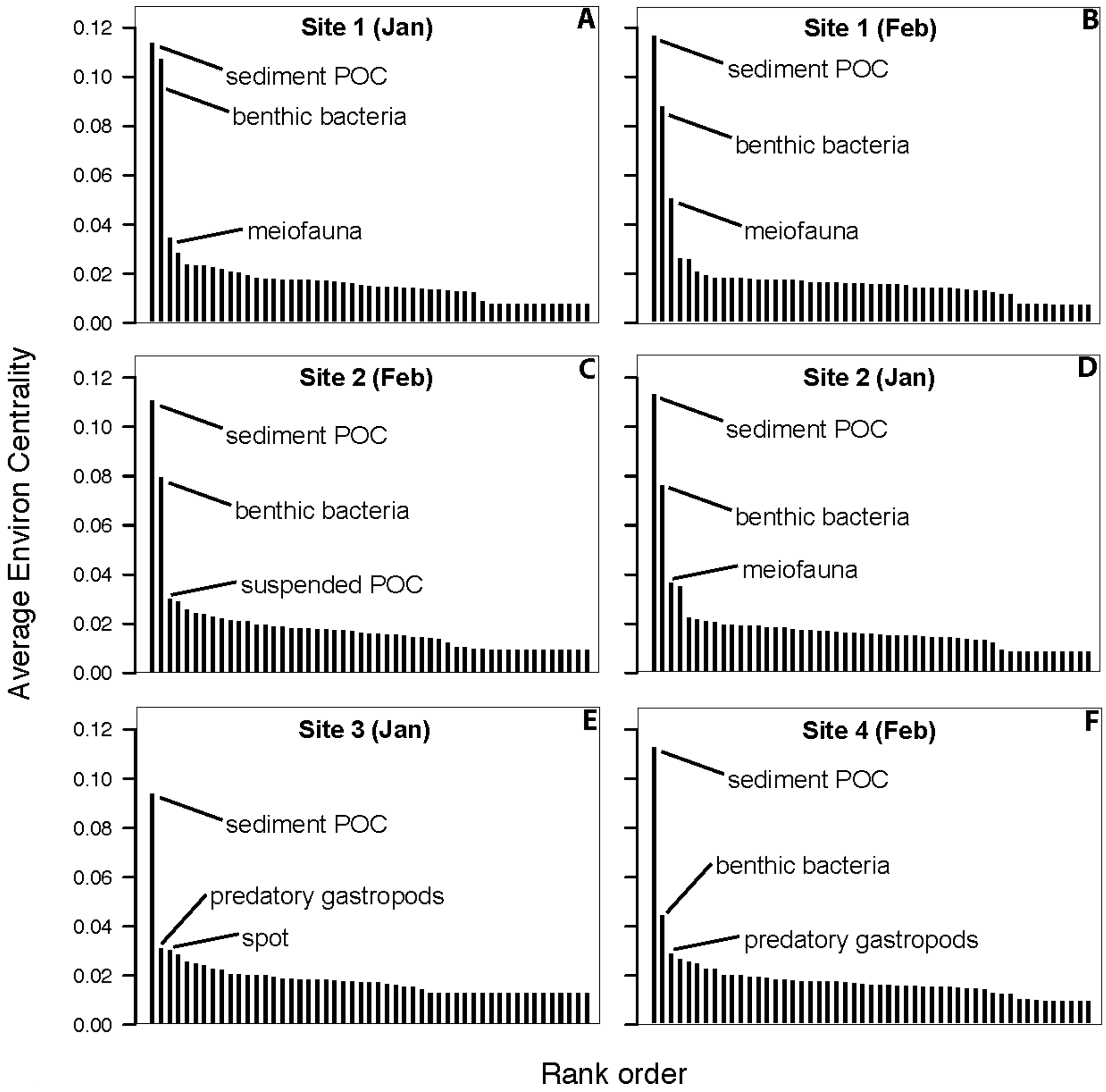

840